\documentclass[,final            % use final for the camera ready runs
  ]
  {aipproc}

\layoutstyle{6x9}
\usepackage{graphicx,wrapfig,caption}

\begin{document}
\title{Magnetic Field Induced Shear Flow in a Strongly Coupled Complex Plasma}

\classification{52.27.Lw, 52.27.Gr, 47.20.Ft}
\keywords      {Strongly coupled complex plasmas, Magnetic fields, Shear flow}

\author{P. Bandyopadhyay}{
  address={Max-Planck Institute for Extraterrestrial Physics, 
Giessenbachstrasse, Garching-85748, Germany}
}

\author{U. Konopka}{
  address={Max-Planck Institute for Extraterrestrial Physics, 
Giessenbachstrasse, Garching-85748, Germany}
}

\author{K. Jiang}{
  address={Max-Planck Institute for Extraterrestrial Physics, 
Giessenbachstrasse, Garching-85748, Germany}
  ,altaddress={Max-Planck Institute for Extraterrestrial Physics, 
Giessenbachstrasse, Garching-85748, Germany} % additional visiting address
}
\author{G. Morfill}{
  address={Max-Planck Institute for Extraterrestrial Physics, 
Giessenbachstrasse, Garching-85748, Germany}
  ,altaddress={Max-Planck Institute for Extraterrestrial Physics, 
Giessenbachstrasse, Garching-85748, Germany} % additional visiting address
}
\begin{abstract}
We address an experimental observation of shear flow of micron sized dust particles in a strongly coupled complex plasma in presence of a homogeneous magnetic field. Two concentric Aluminum rings of different size are placed on the lower electrode of a radio frequency (rf) parallel plate discharge. The modified local sheath electric field is pointing outward/inward close to the inner/outher ring, respectively. The microparticles, confined by the rings and subject to an ion wind that driven by the local sheath electric field and deflected by an externally applied magnetic field, start flowing in azimuthal direction. Depending upon the rf amplitudes on the electrodes, the dust layers show rotation in opposite direction at the edges of the ring-shaped cloud resulting a strong shear in its center. MD simulations shows a good agreement with the experimental results.
\end{abstract}
\maketitle
\section{Introduction}
Complex plasmas are more and more used to study the behavior of fluids on a nano flows equivalent scale. Shear flows built one of the studied topics. As a class of the interfacial hydrodynamics these are attributed to many natural phenomena . The shear in the complex plasma systems can be induced by a dedicated gas flow or by laser beam particle interaction~\cite{nosenko}. Instabilities in the shear flow between two adjacent layers of the fluids can evolve \cite{aswin}. In this paper we report an experimental observation of a shear flow induced by a novel driving methods, that is based on the deflection of the ion wind by an externally applied magnetic field.  
\section{Experimental Setup, results and discussion}
The experiments are carried out in a cylindrical discharge chamber placed inside a superconducting, magnetic coil \cite{uwe} . Plasma is produced by applying a rf-voltage (13.56~MHz) of opposite polarity across two parallel plate electrodes. MF microspheres of 9.19 $\mu$m diameter are injected into the plasma volume, levitated in between two Aluminum rings a few mm above the lower electrode. The laser illuminated particles are observed from above, through the upper, ITO coated transparent electrode. Experiments are conducted at pressures within $p=20-80$~Pa (Argon), a discharge power of a few Watt and an externally applied, downward pointing magnetic field equivalent of up to 0.5~T.\par
\begin{figure}[ht]
\hspace*{1.0in}
\begin{minipage}[t]{2.5in}
\hspace*{-0.8in}
%\centerline{\hbox{figure1a.jpg, width=0.17\textwidth, angle=0}}
\includegraphics[height=0.25\textheight]{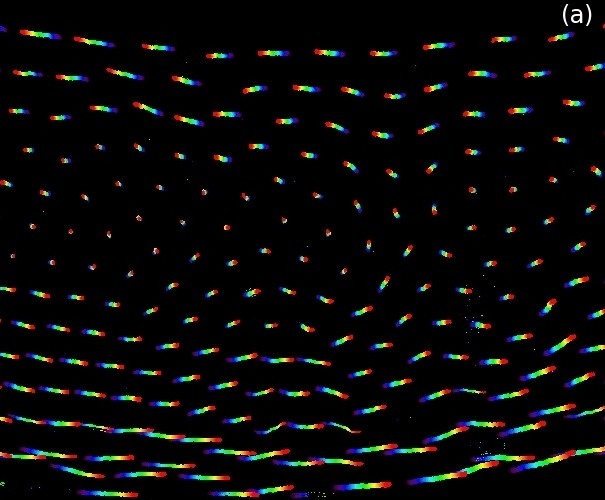}
\end{minipage}
%################################################################### 
\begin{minipage}[t]{2.5in}
%\begin{flushleft}
\hspace*{-0.6in}
\includegraphics[height=0.25\textheight]{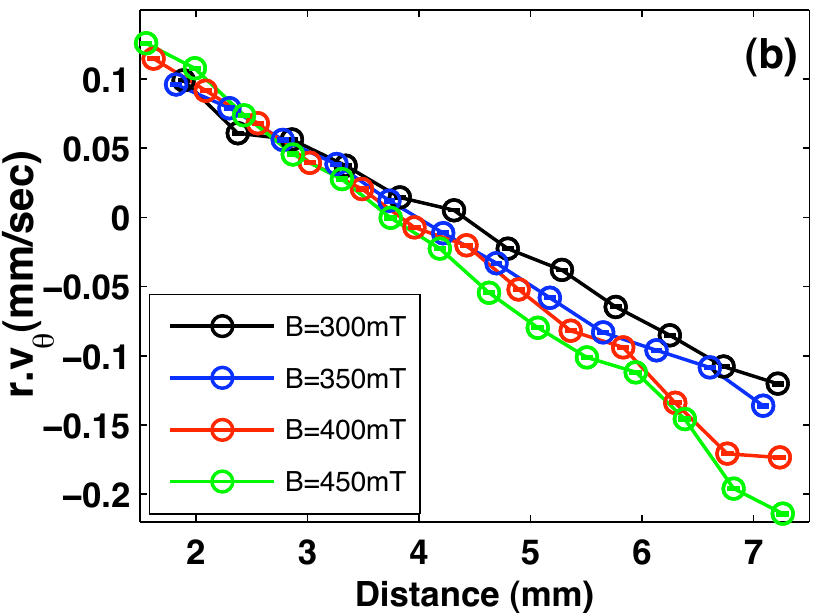}
%\end{flushleft}
\end{minipage}
%################################################################### 
\caption{(a) Experimental observation of shear flow of 9.19 $\mu$m particle at $P=80$Pa, $B=0.4$T at peak to peak electrode voltages $V_{up}=61$V and $V_{down}=35$V. (b) The variation of shear velocity with magnetic fields at $P=80$Pa}
\label{fig:figure1and2}
\end{figure}

Under the influence of the magnetic field, the particles rotate either in clock wise or counter clock wise direction, depending on the applied rf-voltages \cite{uwe}. At particular voltages, the dust layers show rotation in opposite direction at the inner and outer edges of the ring-shaped cloud (figure \ref{fig:figure1and2}(a)).  At this parameters, the radial confining sheath electric fields near the outer edge of the inner ring and the inner edge of the outer ring are pointing in opposite directions. The local $\vec{E} \times \vec{B}$ deflected ion wind is driving the shear flow leading to a shear flow pattern as given in figure \ref{fig:figure1and2}(b). 
%##################################################################
%figure(1)
%###################################################################

The velocity of the shear flow has a strong dependency on magnetic field strength and neutral gas pressure. The linear slope of the velocity profile grows with an increase of magnetic field strength as shown in figure \ref{fig:figure1and2}(b) as well as with an increasing pressure (not shown in the figure).  The increase of velocity with the magnetic field strength is well understood \cite{uwe}. The effect of the pressure is more indirect. With increasing pressure the neutral friction is increasing and thus the shear velocity should decrease. But with increasing neutral pressure (and thus the plasma density) the plasma sheath thickness is decreasing and the confining electric fields steepening. The effect of the confinement strength on the shear flow velocity overcompensates the effect of the friction force. MD simulations confirm the observations.\par
Investigation of magnetic field induced shear flow in a strongly coupled complex plasmas has been reported for the first time. The shear flow, appears at a particular peak to peak electrode voltages, strongly depends on magnetic fields and the neutral gas pressure. The result from MD simulations agree quite well with experimental observations.
%\begin{theacknowledgments}
%The authors acknowledge Richard for his kind help during the experiments.
%\end{theacknowledgments}
\bibliographystyle{aipproc}   % if natbib is available


%%%%%%%%%%%%%%%%%%%%%%%%%%%%%%%%%%%%%%%%%%%
%% The following lines show an example how to produce a bibliography
%% without the help of the BibTeX program. This could be used instead
%% of the above.
%%%%%%%%%%%%%%%%%%%%%%%%%%%%%%%%%%%%%%%%%%%

\end{document}